\documentclass[epj]{svjour}
\usepackage{graphicx,amsmath,amssymb,bm}
\pdfoutput=1

\newcommand{\fmi}{\, \rm{fm}^{-1}}

\newcommand{\mev}{\, \rm{MeV}}
\newcommand{\gevi}{\,\mathrm{GeV}^{-1}}

\begin{document}

\title{Symmetry energy, neutron skin, and neutron star radius 
from chiral effective field theory interactions}

\author{K.\ Hebeler\inst{1,2} \and A.\ Schwenk\inst{2,1}}

\institute{Institut f\"ur Kernphysik,
Technische Universit\"at Darmstadt, 64289 Darmstadt, Germany
\and ExtreMe Matter Institute EMMI,
GSI Helmholtzzentrum f\"ur Schwerionenforschung GmbH, 64291 Darmstadt, Germany}

\abstract{We discuss neutron matter calculations based on chiral effective 
field theory interactions and their predictions for the symmetry energy, 
the neutron skin of $^{208}$Pb, and for the radius of neutron stars.}

\titlerunning{Symmetry energy, neutron skin, and neutron star radius 
from chiral EFT interactions}

\maketitle

\section{Introduction}

\begin{figure}[t]
\begin{center}
\includegraphics[trim=23mm 18mm 149mm 29mm,width=0.425\textwidth,clip=]%
{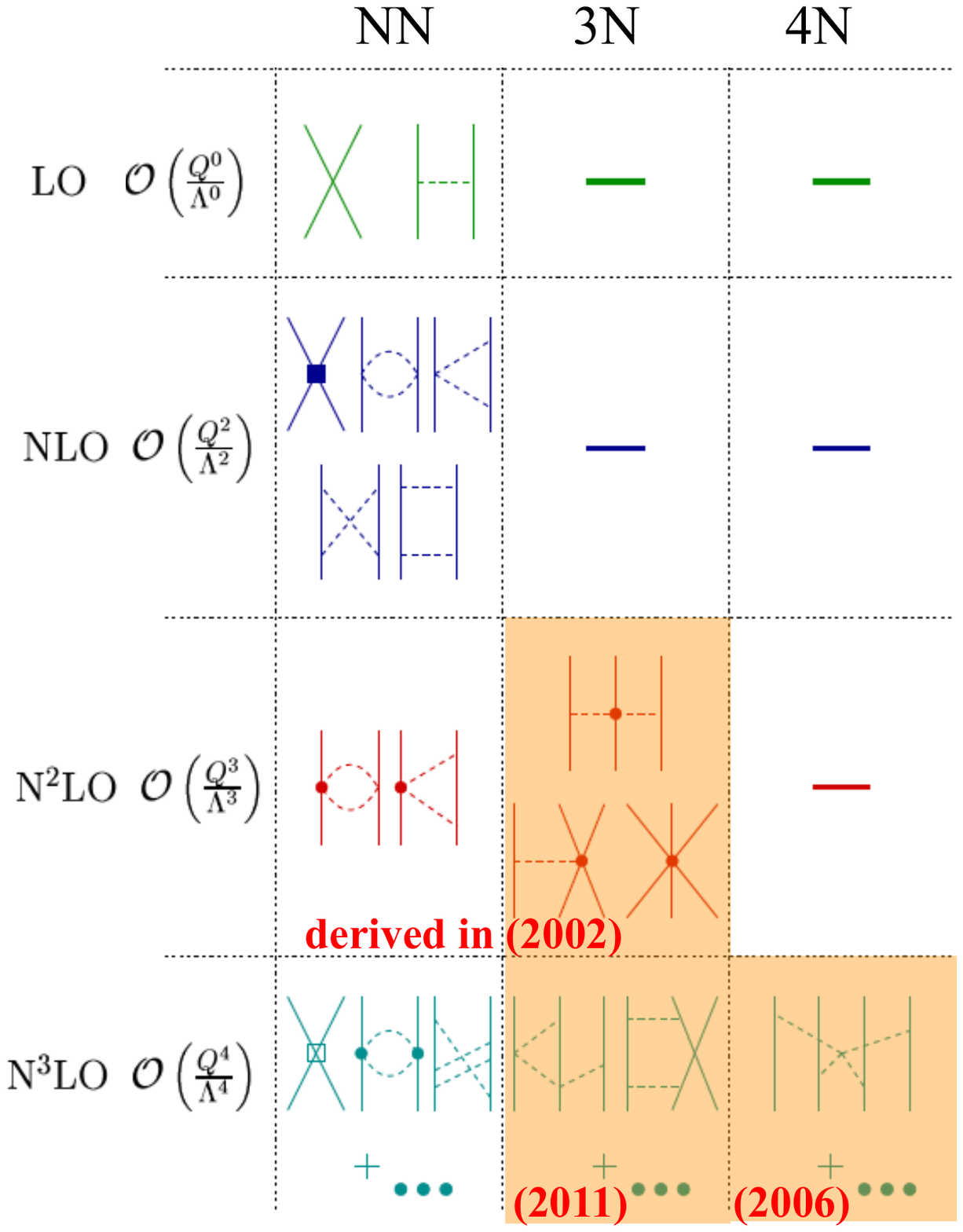}
\end{center}
\caption{Chiral EFT for nuclear forces, where the different contributions
at successive orders are shown diagrammatically~\cite{RMP,EMRept}. 
Many-body forces are highlighted in orange including the year they
were derived. All N$^3$LO 3N and 4N forces are predicted
parameter-free.\label{fig:chiralEFT}}
\end{figure}

Chiral effective field theory (EFT) leads to a systematic expansion
for nuclear forces~\cite{RMP,EMRept}, as shown in Fig.~\ref{fig:chiralEFT},
which provides a powerful approach to three-nucleon (3N)
interactions~\cite{RMP3N} and enables controlled calculations with
theoretical error estimates. This is especially important for exotic
nuclei and neutron-rich matter under extreme conditions in astrophysics.

Neutron matter constitutes a unique system for chiral EFT, because all
3N and four-neutron (4N) interactions are predicted to
next-to-next-to-next-to-leading order (N$^3$LO) without free
parameters~\cite{nm,N3LO,longN3LO}. In addition, neutron matter is a
simpler system, in which one can test the chiral EFT power counting
and the size of many-body forces for densities relevant to nuclei.

\begin{figure*}[t]
\begin{center}
\includegraphics[width=0.425\textwidth,clip=]{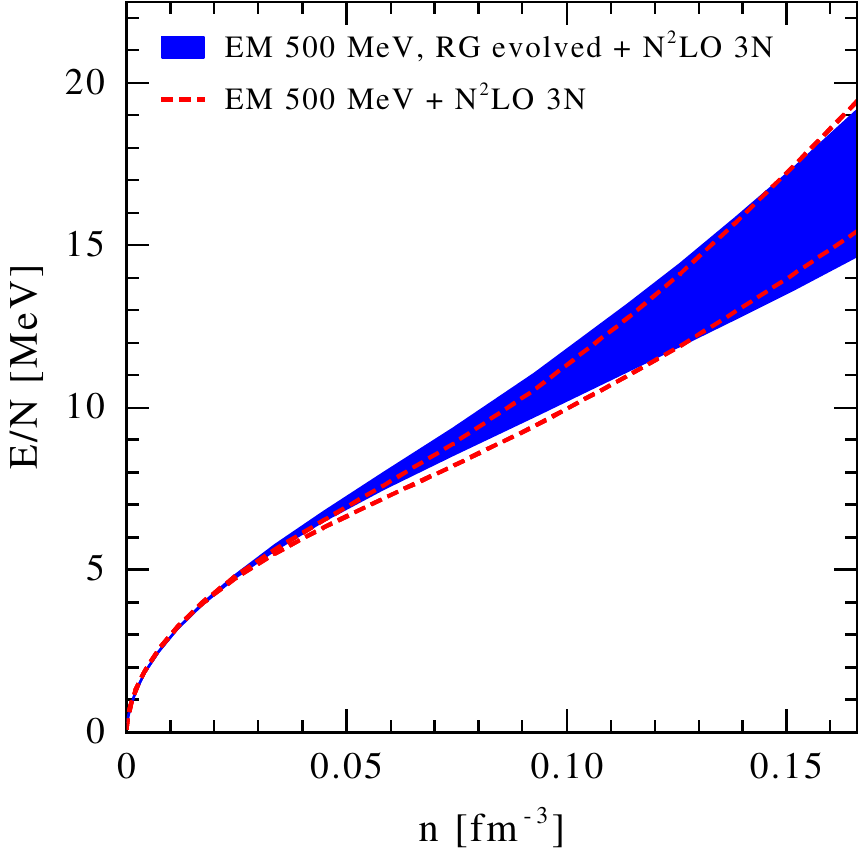}
\hspace{15mm}
\includegraphics[width=0.425\textwidth,clip=]{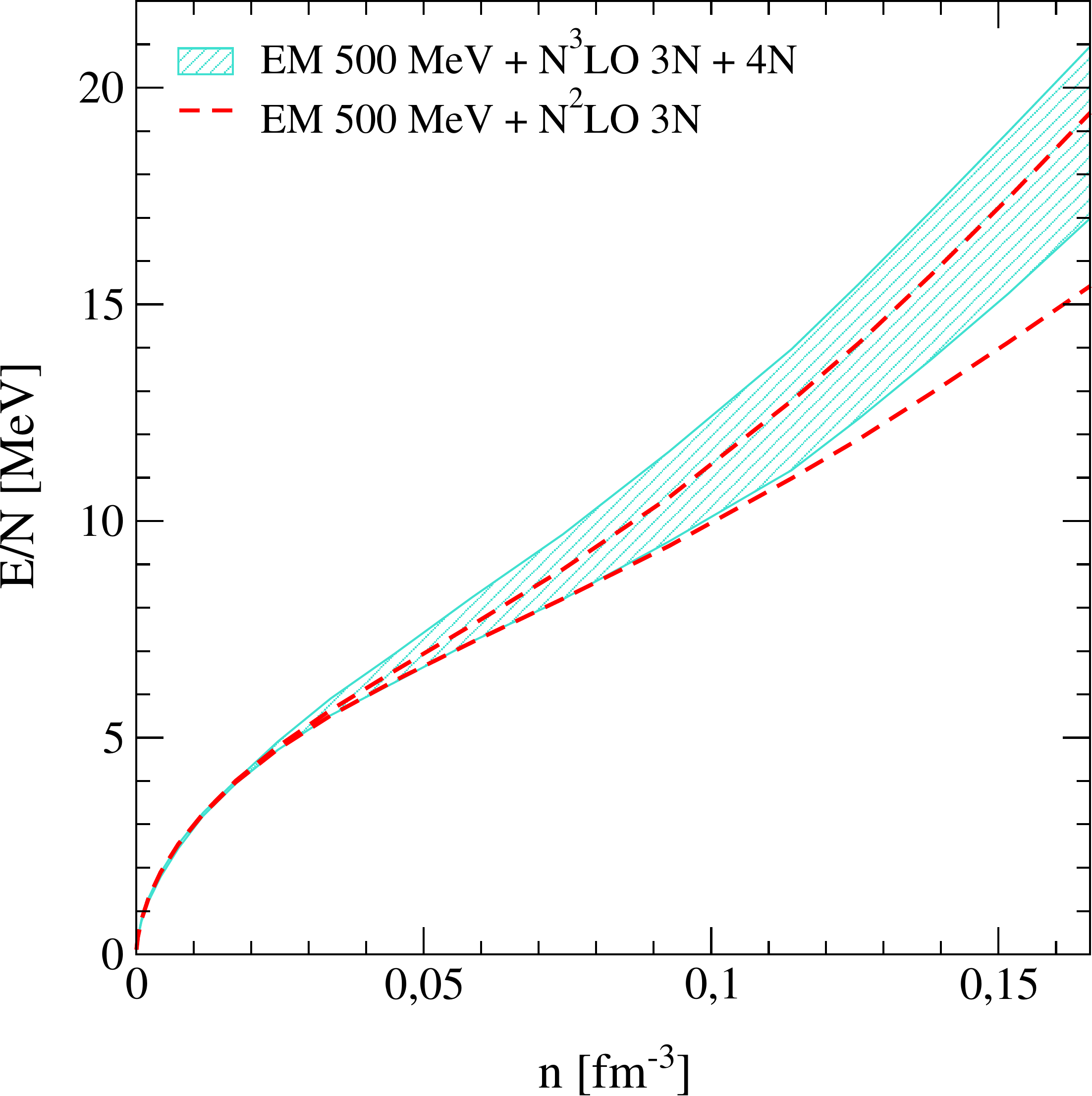}
\end{center}
\caption{Energy per particle $E/N$ of neutron matter as a function of 
density $n$. The dashed red lines show the energy range~\cite{nstar_long}
based on the EM 500 MeV N$^3$LO NN potential of Ref.~\cite{EM} including
N$^2$LO 3N interactions. The blue band in the left panel shows the
corresponding results after RG-evolution of the NN
potential~\cite{nm,nstar_long}. The shaded cyan band in the right
panel shows the results including all 3N and 4N contributions to
N$^3$LO (without RG evolution)~\cite{N3LO,longN3LO}.\label{fig:EN1}}
\end{figure*}

Generally, nuclear forces are not observable and depend on a
resolution scale $\Lambda$, so that the nuclear Hamiltonian is given by
\begin{equation}
H(\Lambda) = T(\Lambda) + V_{\rm NN}(\Lambda) + V_{\rm 3N}(\Lambda) 
+ V_{\rm 4N}(\Lambda) \ldots \,.
\label{eq:Hamiltonian}
\end{equation}
As shown in Fig.~\ref{fig:chiralEFT}, at a given order, nuclear forces
include contributions from one- or multi-pion exchanges, which
constitute the long-range parts, and from contact interactions, whose
scale-dependent short-range couplings are fit to low-energy data for
each $\Lambda$. There are natural sizes to many-body-force
contributions that are made manifest in the EFT power counting and
which explain the phenomenological hierarchy of many-body forces,
$V_{\rm NN}(\Lambda) > V_{\rm 3N}(\Lambda) > V_{\rm 4N}(\Lambda)$~\cite{RMP}.

The renormalization group (RG) is a powerful tool to systematically
change the resolution scale $\Lambda$, while preserving low-energy
observables. The evolution to lower scales facilitates the solution of
the nuclear many-body problem due to a decoupling of low and high
momenta in the Hamiltonian~\cite{PPNP,Kai_review}. In general, RG
transformations change all terms of the
Hamiltonian~(\ref{eq:Hamiltonian}). Recently, it has become possible
to evolve 3N forces in momentum space~\cite{3N_evolution_mom}, with
first applications for neutron matter~\cite{nm_evolved}.

\section{Neutron matter based on chiral EFT interactions}

At low resolution scales, the energy of nucleonic matter can be
calculated with theoretical uncertainties in a perturbative expansion
around the Hartree-Fock energy~\cite{nm,nucmatt1,nucmatt2,asymmatt}.
Figure~\ref{fig:EN1} shows the energy per particle of neutron matter
up to saturation density $n_0 = 0.16 \, {\rm fm}^{-3}$. The results
are based on the EM 500\,MeV N$^3$LO NN potential of Ref.~\cite{EM}.
The blue band in the left panel shows the energy per particle after
RG-evolution of the NN potential to a low-momentum scale $\Lambda =
2.0 \fmi$ and including N$^2$LO 3N interactions~\cite{nm,nstar_long}.
The same Hamiltonians fit only to light nuclei predict nuclear matter
saturation with theoretical uncertainties~\cite{nucmatt2}. At these
scales, NN interactions derived from different NN potentials are very
similar~\cite{Vlowk}. This universality can be attributed to the
common long-range pion physics and the phase-shift equivalence of
high-precision NN interactions. Consequently, the results for the
energy per particle of neutron matter starting from different NN
interactions are also very similar at these resolution scales. The
width of the blue band is dominated by the uncertainties of the $c_1$
and $c_3$ couplings in 3N forces~\cite{nm}. Because the leading chiral
3N forces are of long-range character in neutron matter, they are
expected to be to a good approximation invariant under the RG
evolution. Therefore, we use the N$^2$LO 3N interactions determined by
$c_1$ and $c_3$ also at low-resolution scales. As a comparison, the
dashed red lines show the results based on the unevolved NN
potential. The remarkable overlap indicates that neutron matter is, to
a good approximation, perturbative for chiral NN interactions with
$\Lambda \lesssim 500 \, {\rm MeV}$ (see Ref.~\cite{longN3LO} for
details). This has been benchmarked by first Quantum Monte Carlo (QMC)
calculations with local chiral EFT
interactions~\cite{Gezerlis,NTSE}. In addition, there are calculations
of neutron matter using in-medium chiral perturbation theory
approaches with similar results~\cite{Weise,Ulf}.

Figure~\ref{fig:EN2} shows the first complete N$^3$LO calculation of
the neutron matter energy, which includes all NN, 3N and 4N
interactions to N$^3$LO~\cite{N3LO,longN3LO}. The energy range is
based on different NN potentials, a variation of the couplings $c_1 =
-(0.75 - 1.13)\gevi$, $c_3 = -(4.77 - 5.51) \gevi$~\cite{Krebs}, which
dominates the total uncertainty, a 3N/4N-cutoff variation $\Lambda = 2
- 2.5 \fmi$, and the uncertainty in the many-body calculation. The
N$^3$LO range is in very good agreement with NLO lattice
results~\cite{NLOlattice} and QMC simulations~\cite{GC} at very low
densities (see also the inset), where the properties are determined by
the large scattering length and effective range~\cite{dEFT}. We also
find a very good agreement with other ab initio calculations of
neutron matter based on the Argonne NN and Urbana 3N potentials: In
Fig.~\ref{fig:EN2}, we compare our N$^3$LO results with variational
calculations (APR)~\cite{APR}, which are within the N$^3$LO band, but
do not provide theoretical uncertainties. In addition, we compare the
density dependence with results from Auxiliary Field Diffusion MC
(AFDMC) calculations (GCR)~\cite{GCR} based on nuclear force models
adjusted to an energy difference of $32 \mev$ between neutron matter
and the empirical saturation point.

In the right panel of Fig.~\ref{fig:EN1}, we compare the N$^3$LO
energy obtained from the EM 500\,MeV N$^3$LO NN potential to the
results that include only N$^2$LO 3N interactions (dashed red
lines). Note that it will be important to study the EFT convergence of
3N forces from N$^2$LO to N$^3$LO in more detail, as we find
relatively large individual 3N contributions at N$^3$LO (see
Refs.~\cite{N3LO,longN3LO} for details), and also the low-energy
couplings $c_i$ in NN and 3N interactions depend on the chiral
order~\cite{Krebs}. As a result, the width of the energy bands based
on 3N forces at N$^2$LO and N$^3$LO are comparable at higher
densities.
 
\section{From the neutron matter equation of state 
to the symmetry energy and neutron skin}

\begin{figure}[t]
\begin{center}
\includegraphics[width=0.425\textwidth,clip=]{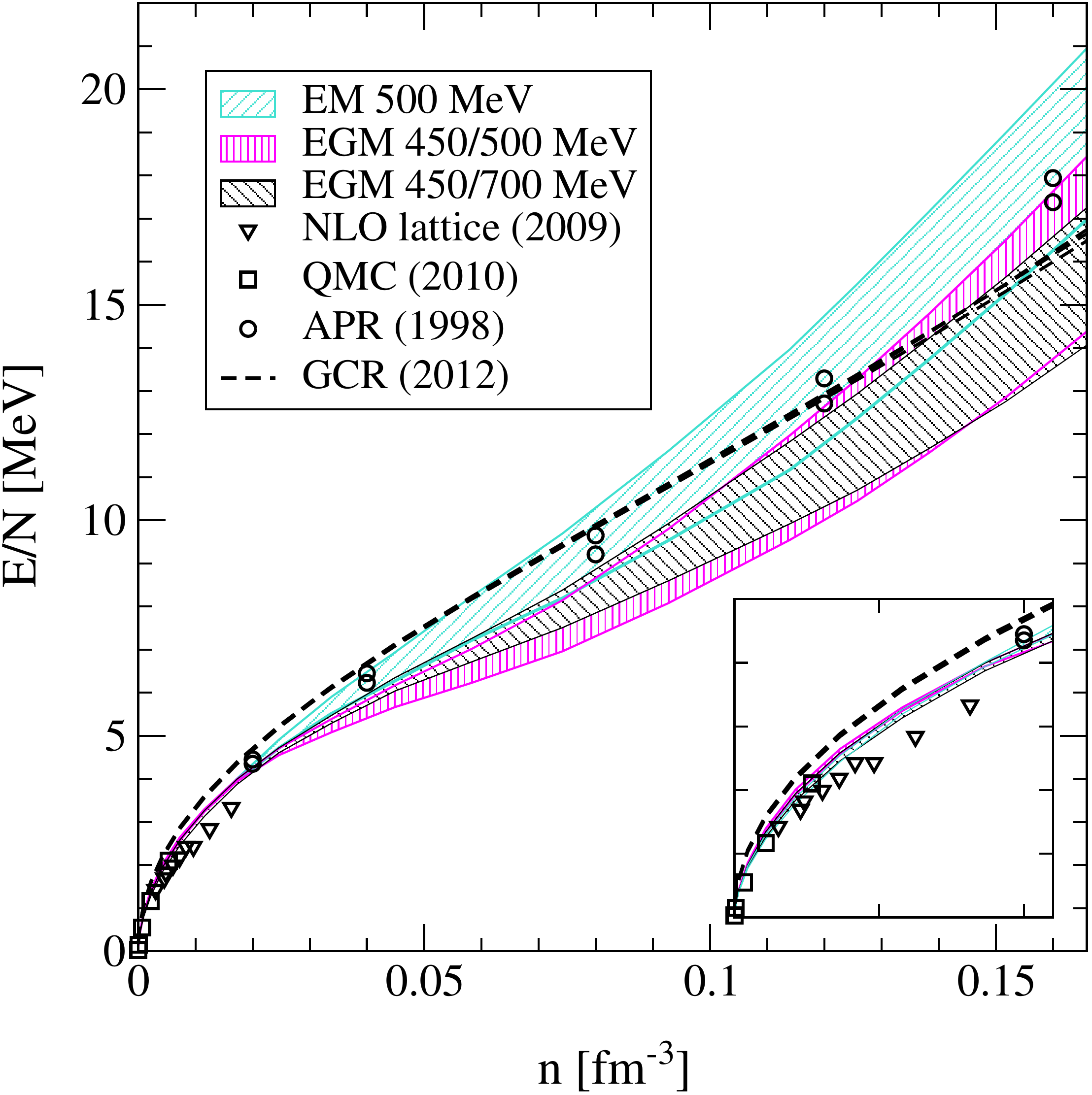}
\end{center}
\caption{Neutron matter energy per particle as a function of density 
including NN, 3N and 4N forces at N$^3$LO. The three overlapping bands
are labeled by the different NN potentials and include uncertainty
estimates due to the many-body calculation, the low-energy $c_i$
constants, and by varying the 3N/4N cutoffs (see
Refs.~\cite{N3LO,longN3LO} for details). For comparison, results are
shown at low densities (see also the inset) from NLO
lattice~\cite{NLOlattice} and Quantum Monte Carlo (QMC)
simulations~\cite{GC}, and at nuclear densities from variational
(APR)~\cite{APR} and Auxiliary Field Diffusion MC calculations
(GCR)~\cite{GCR} based on 3N potentials adjusted to nuclear matter
properties.\label{fig:EN2}}
\end{figure}

We extend our ab initio results for neutron matter to matter with a
finite proton fraction $x= n_p/n$ by using an empirical
parametrization~\cite{nstar_long} of the energy per particle $E/A$
that includes kinetic energy plus interaction energy, which is expanded
in the Fermi momentum and is quadratic in the neutron excess $1-2x$,
\begin{eqnarray}
\frac{E/A(\bar{n},x)}{T_0} &=&
\frac{3}{5} \left[ x^{5/3} + (1-x)^{5/3} \right] (2 \bar{n})^{2/3} 
\nonumber \\
&&- \left[ ( 2 \alpha - 4 \alpha_L) x (1 - x) + \alpha_L \right] \bar{n} \nonumber \\[1mm]
&& + \left[ ( 2 \eta - 4 \eta_L) x (1 - x) + \eta_L \right]
\bar{n}^{4/3} \,. \label{Eskyrme}
\end{eqnarray}
Here $\bar{n} = n/n_0$ is the baryon density in units of the saturation
density and $T_0 = (3 \pi^2 n_0/2)^{2/3} \hbar^2/(2m) = 36.84 \,
\rm{MeV}$ is the Fermi energy of symmetric nuclear matter at
saturation density. The corresponding expression for the pressure follows
from $P=n^2\partial (E/A)/ \partial n$. The empirical
para-metrization interpolates between the properties of neutron matter
and symmetric matter and was recently benchmarked against ab initio
calculations of asymmetric matter with very good
agreement~\cite{asymmatt}.

The parameters $\alpha, \eta$ and $\alpha_L, \eta_L$ can be determined
from the empirical saturation properties of symmetric matter ($x=1/2$)
and from the microscopic calculations of neutron matter ($x=0$),
respectively. The empirical saturation point $E/A(\bar{n}=1, x=1/2) =
-16 \, {\rm MeV}$ and $P(\bar{n}=1, x=1/2) = 0$ gives $\alpha = 5.87$,
$\eta = 3.81$ with a reasonable incompressibility $K = 236 \,
\rm{MeV}$, where the precise value of $K$ could be adjusted by
modifying the exponent $4/3$ in the parametrization~(\ref{Eskyrme}).
However, the predicted range for the symmetry energy and its density
derivative depend very weakly on this choice~\cite{nstar_long}.

\begin{figure}[t]
\begin{center}
\includegraphics[width=0.425\textwidth,clip=]{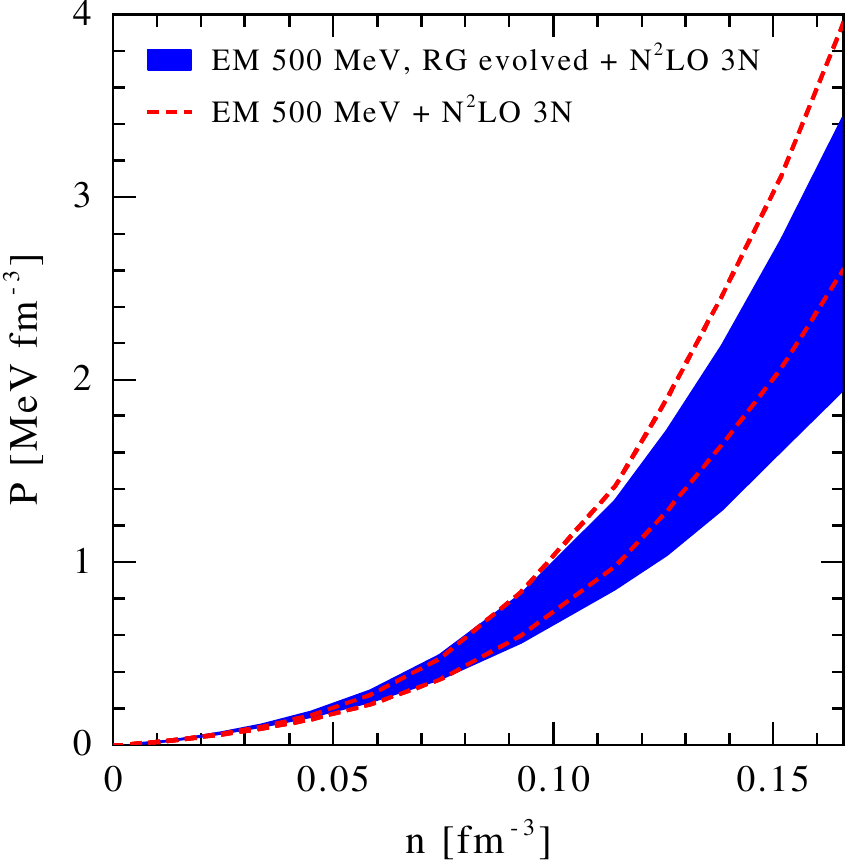}
\end{center}
\caption{Pressure $P$ of neutron matter as a function of density~$n$.
Results are shown for the same NN and 3N interactions as for the energy
in the left panel of Fig.~\ref{fig:EN1}.\label{fig:P}}
\end{figure}

The parameters $\alpha_L, \eta_L$ are extracted from the calculated
bands for the neutron matter energy and pressure. For this, we use the
results based on RG-evolved NN interactions, given by the blue bands
in Fig.~\ref{fig:EN1} for the energy and Fig.~\ref{fig:P} for the
pressure. Due to the improved convergence by the RG, the uncertainties
from the many-body calculation are smaller than for unevolved
interactions. To determine $\alpha_L, \eta_L$, we sample their values
and require that the resulting energy and pressure are within the
uncertainty bands~(see Ref.~\cite{nstar_long} for details).

\begin{figure}[t]
\begin{center}
\includegraphics[width=0.45\textwidth,clip=]{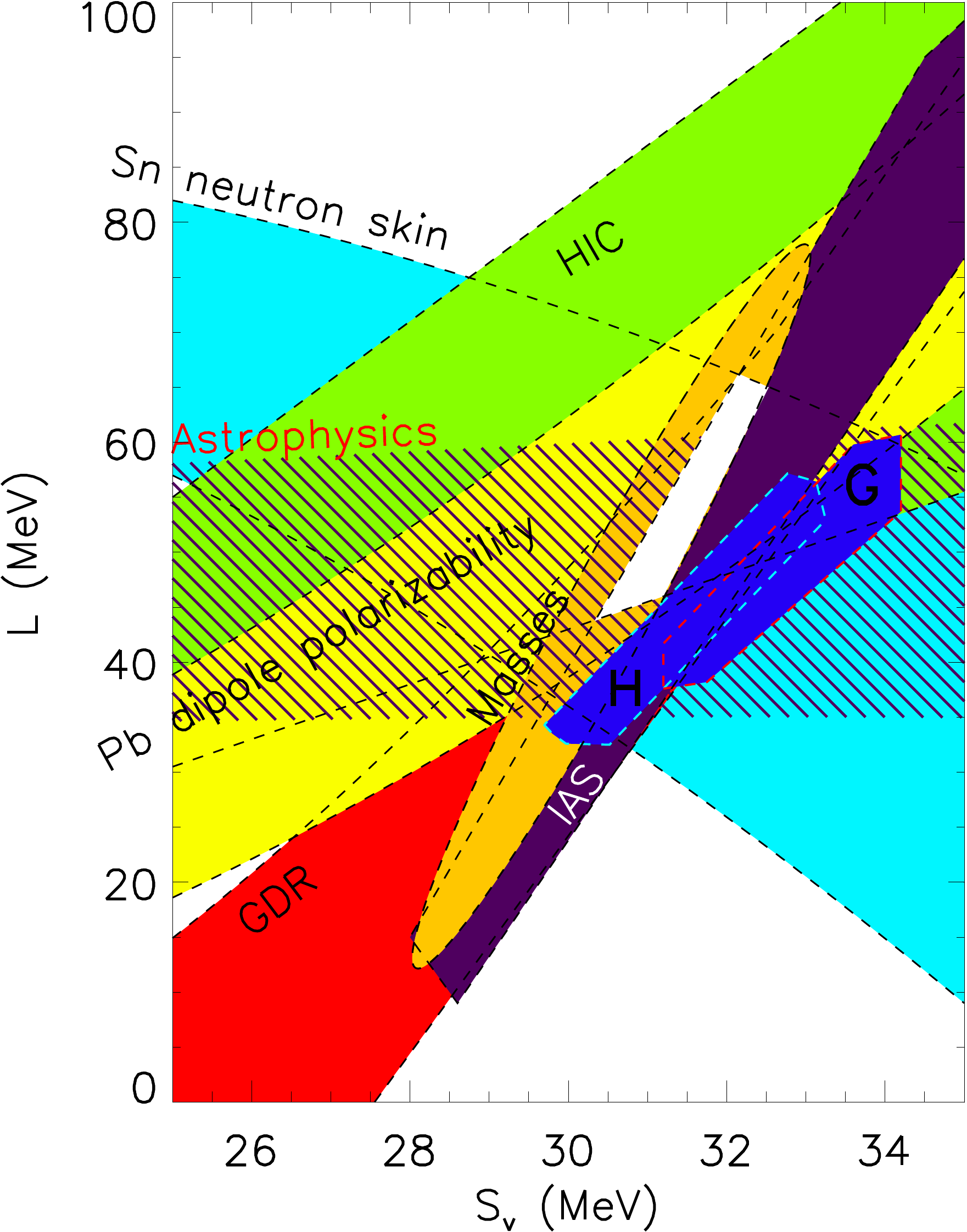}
\end{center}
\caption{Constraints for the symmetry energy $S_v$ and the $L$ parameter
following~\cite{LL}. The blue region labeled ``H'' shows our neutron
matter constraints, ``G'' represents the results of Ref.~\cite{GCR}
adjusted to a range of the symmetry energy. These are compared to bands
based on different empirical extractions (for details see 
Refs.~\cite{nstar_long,SvLnew}). The white area gives the overlap
region of the different empirical ranges.\label{symcor}}
\end{figure}

The parametrization~(\ref{Eskyrme}) allows to predict the symmetry
energy $S_v$ and its density derivative $L$,
\begin{eqnarray}
S_v &=& \frac{1}{8} \frac{\partial^2 \epsilon(\bar{n},x)}{\partial x^2} 
\biggr|_{\bar{n}=1, x=1/2} \\[1mm]
L &=& \frac{3}{8} \frac{\partial^3 \epsilon(\bar{n},x)}{\partial \bar{n} 
\partial x^2} \biggr|_{\bar{n}=1, x=1/2} \,.
\end{eqnarray}
The values for $\alpha_L, \eta_L$ translate into correlated ranges for
$S_v = 29.7 - 33.2 \, {\rm MeV}$ and $L = 32.5 - 57.0 \, {\rm
MeV}$~\cite{nstar_long}. In Fig.~\ref{symcor}, we compare the
$S_v-L$ region predicted by our neutron matter results with bands
extracted from other analyses~\cite{nstar_long,LL,SvLnew}. Strikingly,
the neutron matter results provide the tightest constraints. They
agree well with constraints obtained from energy-density functionals
for nuclear masses (orange band)~\cite{SvLnew,masses} as well as from
the $^{208}$Pb dipole polarizability (yellow
band)~\cite{SvLnew,Tamii,TamiiEPJA,Witek} and from a recent analysis of
isobaric analog states (IAS, purple band)~\cite{Danielewicz13}. In
addition, there is good agreement with studies of the Sn neutron skin
(light blue band)~\cite{Sn}, of isotope diffusion in heavy ion
collisions (HIC, green band)~\cite{HIC}, and of giant dipole
resonances (GDR, red band) \cite{GDR}. For additional details see
Refs.~\cite{LL,SvLnew}. Moreover, there is very good agreement with an
estimate obtained from modeling X-ray bursts and quiescent low-mass
X-ray binaries (``Astrophysics'', shaded region)~\cite{Steiner}.

It is also remarkable how well the $S_v-L$ region predicted by our
neutron matter results agrees with the one obtained from the AFDMC
calculations~\cite{GCR} based on a set of very different Hamiltonians,
the Argonne $v_8'$ NN and Urbana IX 3N potentials (``G'' in
Fig.~\ref{symcor}), where the region is constructed from different 3N
models with a symmetry energy between the NN only and NN plus Urbana
IX 3N result (an intermediate model in this set is given by the GCR
curve in Fig.~\ref{fig:EN2}).

\begin{figure}[t]
\begin{center}
\includegraphics[width=0.45\textwidth,clip=]{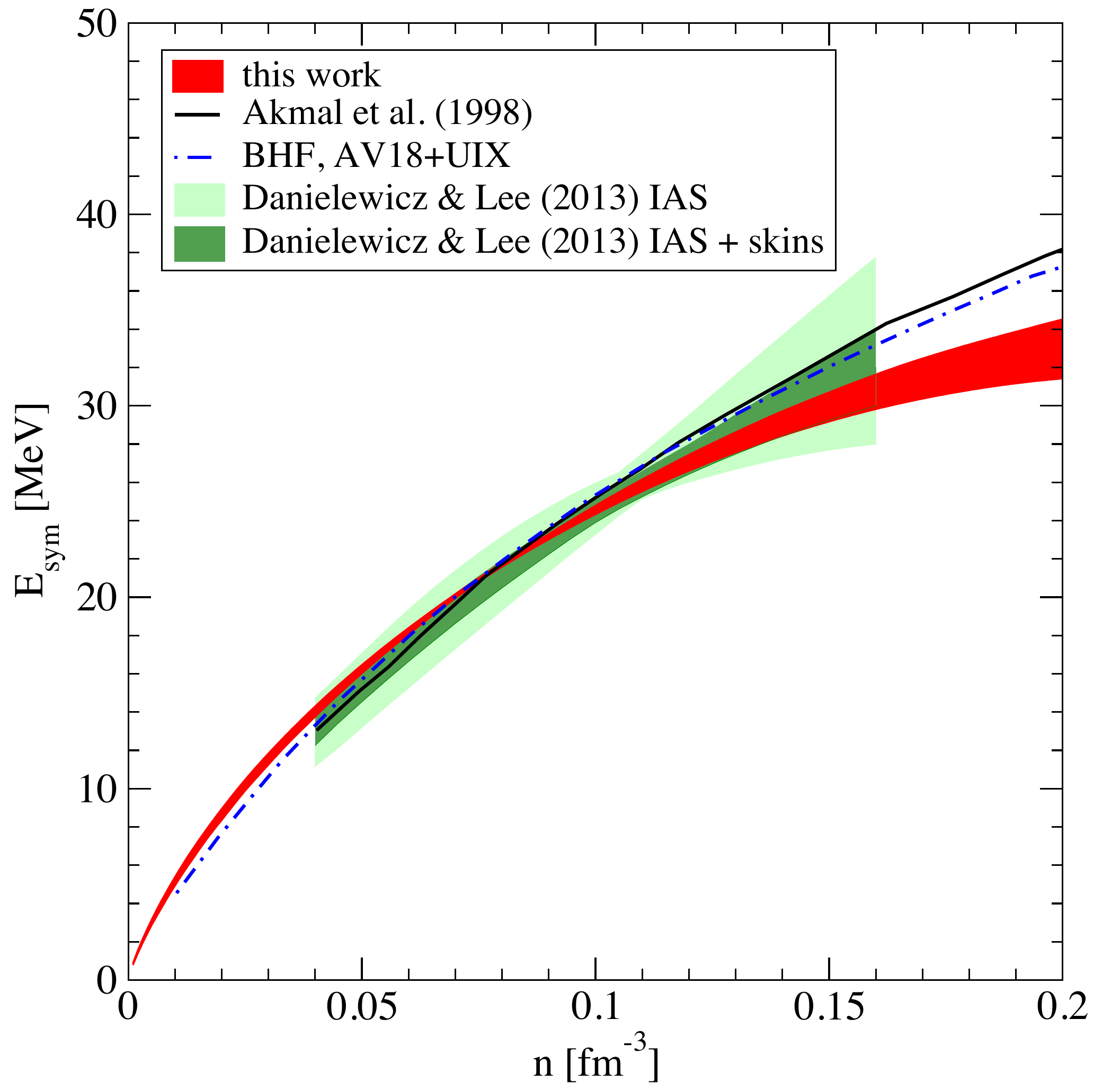}
\end{center}
\caption{$E_\text{sym}$ as a function of density obtained
from ab initio calculations of asymmetric matter~\cite{asymmatt}.
In comparison, we give $E_\text{sym}$ obtained from microscopic
calculations performed with a variational approach (Akmal et
al.~(1998))~\cite{APR} and at the Brueckner-Hartree-Fock level
(BHF)~\cite{Taranto13} based on the Argonne $v_{18}$ NN and Urbana UIX
3N potentials (with parameters adjusted to the empirical saturation
point). The band over the density range $n=0.04-0.16 \, {\rm fm}^{-3}$
is based on a recent analysis of isobaric analog states (IAS) and
including the constraints from neutron skins (IAS +
skins)~\cite{Danielewicz13}.\label{fig:Esym_vs_n}}
\end{figure}

Recently, the symmetry energy was studied in ab initio calculations of
asymmetric matter based on N$^3$LO NN and N$^2$LO 3N
interactions~\cite{asymmatt}. The energy of asymmetric matter was
found to compare very well with a quadratic expansion even for
neutron-rich conditions, which was then used to extract the quadratic
symmetry-energy term $E_{\rm sym}$. In contrast to other calculations,
the results are based on 3N forces fit only to light nuclei, without
adjustments to empirical nuclear matter properties. The results for
$E_{\rm sym}$ are compared in Fig.~\ref{fig:Esym_vs_n} with constraints
from a recent analysis of isobaric analog states (IAS) and including
the constraints from neutron skins (IAS + skins)~\cite{Danielewicz13},
showing a remarkable agreement over the entire density range.
Note that compared to extracting the symmetry energy from neutron
matter calculations using the parametrization~(\ref{Eskyrme}), the
uncertainty is reduced due to the explicit information from 
asymmetric matter.

The neutron skin probes the properties of neutron matter, as a higher
pressure at typical nuclear densities implies larger neutron
skins~\cite{Brown}. Using these correlations~\cite{Brown} (and
including a study based on the liquid droplet model), our neutron
matter results of the blue bands in Figs.~\ref{fig:EN1}
and~\ref{fig:P} predict the neutron skin of $^{208}$Pb to $0.17 \pm
0.03 \, {\rm fm}$~\cite{nstar}, which is in excellent agreement with
the extraction of $0.156\substack{+0.025 \\ -0.021} \, {\rm fm}$ from
the dipole polarizability~\cite{Tamii}. The theoretical uncertainty is
also smaller than the target goal of a new PREX measurement using
parity violating electron scattering at JLAB~\cite{PREX}. Moreover,
including properties of doubly-magic nuclei as constraints, in
addition to low-density neutron matter results, leads to even tighter
predictions for the neutron skins of $^{208}$Pb and $^{48}$Ca to be
$0.182 \pm 0.010 \, {\rm fm}$ and $0.173 \pm 0.005 \, {\rm fm}$,
respectively~\cite{skins}.

\section{Constraints on neutron star radii}

\begin{figure}[t]
\begin{center}
\includegraphics[width=0.45\textwidth,clip=]{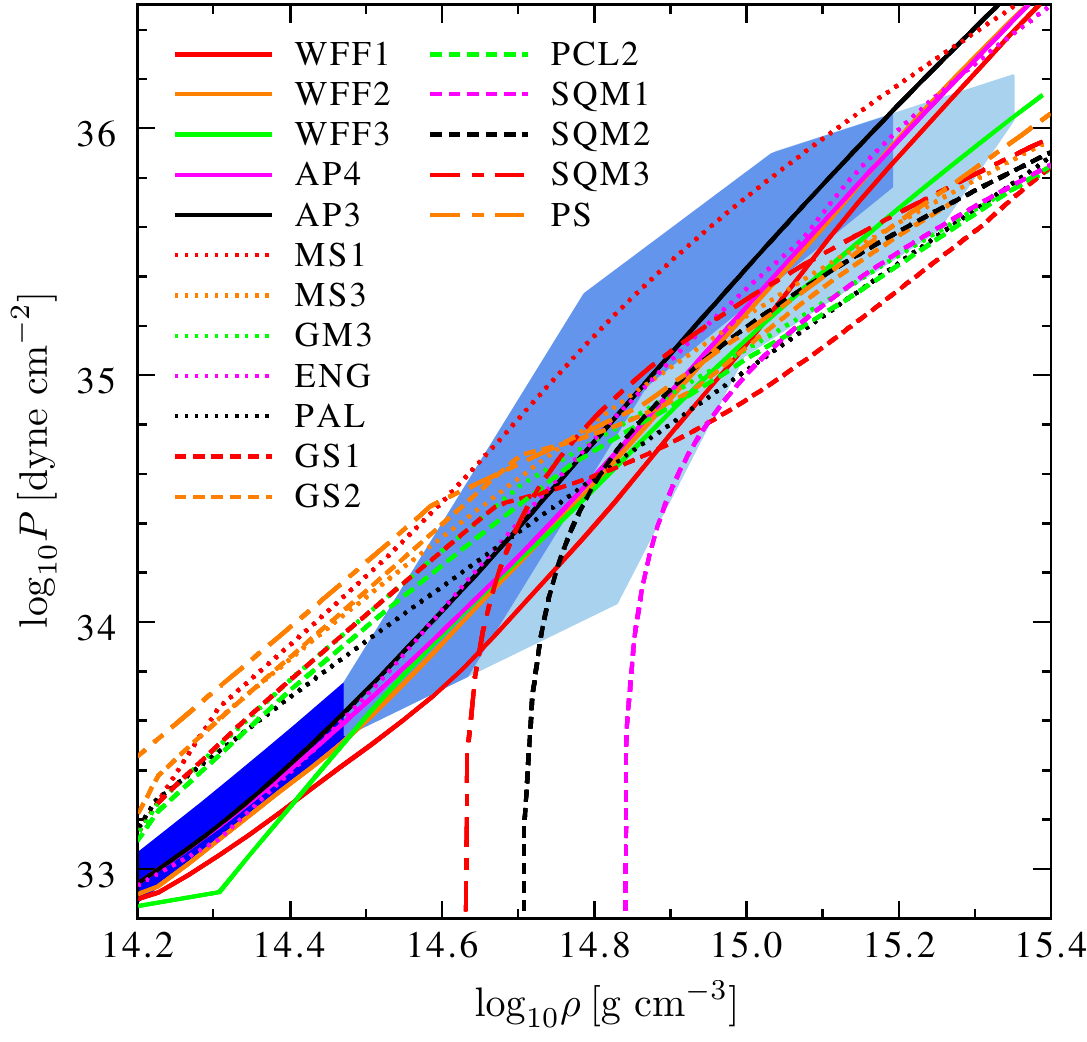}
\end{center}
\caption{Constraints for the pressure $P$ of neutron star matter as a
function of mass density $\rho$ compared to EOSs commonly used to model
neutron stars~\cite{LP}. The lighter blue band is the envelope of the
general polytropic extensions that are causal and support a neutron
star of mass $\widehat{M} = 1.97 \, M_\odot$ and the darker blue band
at high densities corresponds to $\widehat{M} = 2.4 \, 
M_\odot$~\cite{nstar_long}.\label{fig:prho_compare}}
\end{figure}

We also use the parametrization~(\ref{Eskyrme}) to extend the neutron
matter results to neutron star matter in beta equilibrium. The proton
fraction in beta equilibrium is determined by minimizing the total
energy per particle with respect to $x$ at a given density including
the contributions from electrons and from the rest mass of the nucleons,
\begin{equation}
\frac{\partial E/A(\bar{n},x)}{\partial x} + \mu_e(\bar{n},x) 
- (m_n - m_p) c^2 = 0 \,.
\end{equation}
Based on the neutron matter bands, this leads to a proton fraction of
$4\% - 5.3\%$ at saturation density and a crust-core transition
density at $n \approx n_0/2$~\cite{nstar_long}. To describe the
equation of state (EOS) of neutron star matter, we use the BPS outer
crust EOS for densities below $n_0/2$~\cite{BPS,Vautherin}. Note that
without 3N forces, the calculated EOSs would not match on to a
standard crust EOS~\cite{nstar}.

Because the central densities of neutron stars can significantly
exceed the regime of our neutron matter calculations, we extend the
EOSs for $n > 1.1 \, n_0$ by employing general piecewise polytropic
extensions~\cite{nstar,nstar_long}. This strategy generates a very
large number of EOSs, allows for soft regions and constitutes a
complete set of possible EOSs at high densities, independent of the
assumptions on the interactions and constituents of matter at high
densities. We solve the Tolman-Oppenheimer-Volkov equations for each
of these EOSs and retain only those that 1)~remain causal for all
relevant densities, and 2)~are able to support a neutron star mass $M
= \widehat{M}$, the mass of the heaviest neutron star observed or
potential heavier candidates.

\begin{figure}[t]
\begin{center}
\includegraphics[width=0.425\textwidth,clip=]{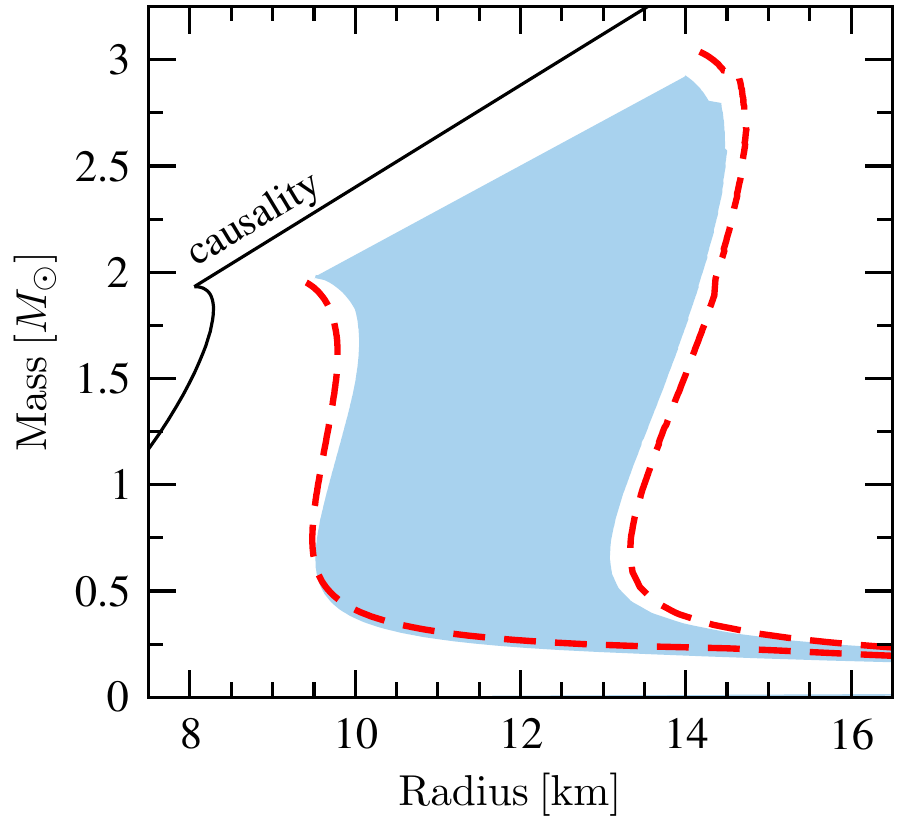}
\end{center}
\caption{Neutron star mass-radius constraints based on the uncertainty
band for the EOSs of Fig.~\ref{fig:prho_compare} for $\widehat{M} = 
1.97 \, M_{\odot}$. The blue region gives the radius constraints based
on the neutron matter results with RG-evolved interactions, the red 
dashed lines based on unevolved interactions, as in the left panel 
of Fig.~\ref{fig:EN1}.\label{fig:MvsR}}
\end{figure}

In Fig.~\ref{fig:prho_compare}, we compare the resulting EOS
uncertainty bands with a representative set of EOSs used in the
literature~\cite{LP}. The low-density pressure sets the scale, and
then the lighter blue band correspond to the mass constraint
$\widehat{M} = 1.97 \, M_\odot$, the central value of the
two-solar-mass neutron star measured by Shapiro delay~\cite{Demorest}
and the lower $1\sigma$ mass of the recently observed most massive
neutron star from radio timing observations~\cite{Antoniadis}, whereas
the darker blue band corresponds to $\widehat{M} = 2.4 \, M_\odot$, a
fictitious heavier neutron star. Figure~\ref{fig:prho_compare}
demonstrates that chiral EFT interactions provide strong constraints,
ruling out many model EOSs at low densities and, combined with the
astrophysics constraint of a heavy neutron star, at high densities as
well. Table~\ref{tab:core_dens} shows that these constraints imply
that a $1.4 \, (1.97) \, M_\odot$ neutron star does not exceed
densities beyond $4.4 \, (7.6) \, n_0$, which corresponds to a Fermi
momentum of only $550 \, (660) \, {\rm MeV}$.

\begin{table}[t!]
\begin{center}
\begin{tabular}{l|cc|cc}
& \multicolumn{2}{c|}{$\widehat{M} = 1.97 \, M_{\odot}$} 
& \multicolumn{2}{c}{$\widehat{M} = 2.4 \, M_{\odot}$} \\
& min & max & min & max \\
\hline
$\rho_{c}/\rho_0$ ($1.4 \,M_{\odot}$)  & 1.8 & 4.4 & 1.8 & 2.7 \\
$\rho_{c}/\rho_0$ ($1.97 \,M_{\odot}$) & 2.0 & 7.6 & 2.0 & 3.4 \\
$\rho_{c}/\rho_0$ ($2.4 \,M_{\odot}$)  &     &     & 2.2 & 5.4 \\
\end{tabular}
\end{center}
\caption{Minimal and maximal central densities $\rho_c$ (in units of
the saturation density $\rho_0$).\label{tab:core_dens}}
\end{table}

From the EOS uncertainty bands in Fig.~\ref{fig:prho_compare} we can
directly derive constraints for the radii of neutron stars. In
Fig.~\ref{fig:MvsR}, we present the radius constraints obtained from
the EOS band of Fig.~\ref{fig:prho_compare} for $\widehat{M} = 1.97 \,
M_{\odot}$. We calculated the mass-radius relationships for the
individual EOSs by solving the Tolman-Oppenheimer-Volkov
equations. Based on all these results we constructed an envelope based
on the extreme values (see Ref.~\cite{nstar_long} for details). The
blue region (red dashed lines) in Fig.~\ref{fig:MvsR} show the radius
constraints based on the neutron matter results with RG-evolved
(unevolved) interactions, as in the left panel of Fig.~\ref{fig:EN1}.
For a typical $1.4 \, M_{\odot}$ neutron star, we predict a radius range
of $R=9.7 - 13.9 \, {\rm km}$. Our radius range is also consistent
with astrophysical extractions obtained from modeling X-ray burst
sources (see, e.g., Ref.~\cite{Steiner}), and the neutron matter
constraints have recently been explored for the gravitational wave
signal in neutron-star mergers~\cite{Bauswein}. Finally, all EOSs for
cold matter in beta equilibrium should go through the EOS uncertainty
bands, independent of composition, and we have constructed three
representative EOSs for astrophysical applications~\cite{nstar_long}.

\section{Summary and outlook}

We have shown that the properties of neutron-rich matter at nuclear
densities are well constrained by chiral EFT interactions. This
results in tight constraints for the symmetry energy, the neutron skin
of $^{208}$Pb, and for the radius of neutron stars. The theoretical
uncertainties are dominated by the uncertainties in 3N
forces. Therefore, developments in 3N forces will be important next
steps for nuclei and nucleonic matter. These include their consistent
SRG evolution, improved treatments with N$^3$LO interactions, and by
going to a Delta-full EFT.

Neutron-rich matter is also the focus of rare isotope beam facilities
worldwide, where predictions based on the same nuclear forces can be
tested in exotic nuclei. This goes hand in hand with advances in
many-body methods and studying 3N forces in medium-mass neutron-rich
nuclei, where the frontier of ab initio calculations is presently in
the calcium
region~\cite{Calcium,CCCa,Roth,TITAN,pairing,ISOLDE,Vittorio}. This
promises further interesting developments from ab initio calculations
for the neutron skin of $^{48}$Ca.

\section*{Acknowledgements}

We would like to thank C.\ Drischler, T.\ Kr\"uger, J.\ M.\ Lattimer,
C.\ J.\ Pethick, V.\ Som\`a, and I.\ Tews, who contributed to the
results discussed in this short topical review. This work was supported by
the DFG through Grant SFB 634, by the ERC Grant No.~307986 STRONGINT,
and by the Helmholtz Alliance HA216/EMMI.

\end{document}